\documentstyle[twoside,fleqn,11pt,espcrc1]{article}

\newcommand{\avX}{\mbox{$\langle X\rangle$}}
\newcommand{\avXt}{\mbox{$\langle X(t)\rangle$}}

\title{ 
{\em\small Expanded version of the contribution
 in ``Field Theory and Statistical Mechanics, Rome 10-15 June 2002''}\\[1cm]
 \bf Particle oscillations in external chaotic fields}

\author{E. Torrente-Lujan
\address{Dept.  F\'{\i}sica
Te\'orica C-XI, Univ. Aut\'onoma de Madrid, Cantoblanco, 28049 Madrid, Spain.}
}

\begin{document}

\maketitle

\begin{abstract}
We review here   the  development of the general 
 formalism for the study of fermion
 propagation in the presence of
stochastic media. This formalism  allows 
 the systematic derivation of evolution equations for averaged 
quantities as survival probabilities and higher order distribution moments. 
The formalism equally applies to 
 any finite dimensional Schr\"{o}dinger equation in the presence of a 
stochastic external field. 
New integrodifferential equations
 valid for finite 
correlated processes are obtained for the first time. 
For the particular case of  
 exponentially correlated processes a second order ordinary equation is 
obtained.
As a consequence, the Redfield equation valid
 for Gaussian delta-correlated
noise is rederived in  a simple way: 
it has been obtained directly and as
the zero-order term of an asymptotic expansion in 
 the inverse  of the correlation length.
The formalism, together
with the quantum correlation theorem is applied  to the computation of  
higher moments and correlation functions.
It is shown that equal and unequal time correlators follow similar differential equations.

\end{abstract}


\section{INTRODUCTION}

We review here the general 
 formalism for the study of neutrino propagation in the presence of
stochastic media \cite{XX}. 
This formalism will  allow 
 the systematic derivation of evolution equations for averaged 
quantities as survival probabilities and higher order moments of them. 
New integrodifferential equations
 valid for finite 
correlation processes are obtained for the first time. 
For exponentially correlated processes a second order 
ordinary equation is 
obtained as a consequence.
The Redfield equation valid
 for Gaussian delta-correlated
noise is rederived in  a simple way as a particular case: 
it can be obtained directly or as
the zero-order term of an asymptotic expansion in 
 the inverse  of the correlation length.
 In the context of neutrino oscillations,
it is shown that the presence of matter noise induces the appearance  of 
a effective complex part in the matter density. 
Finally the formalism, together
with the quantum correlation theorem, will be applied 
to the problem of the computation of distribution 
higher moments and non-equal-time correlation functions.

The formalism may be applied to any 
quantum system governed by similar Schr\"{o}dinger equations: equations where
a stochastic function appears multiplicatively in some part of the 
Hamiltonian. For simplicity of notation, in this work we will   always
deal 
 with Hamiltonians which are finite dimensional operators in some 
Hilbert space,  the equations will be obviously of the 
same type  for general systems
described by infinite-dimensional 
Hamiltonians.

\section{THE GENERAL CASE}
 
Let us consider a general system  whose evolution is described by 
the following linear random Schr\"{o}dinger  equation:
\begin{eqnarray}
i\partial_t X &=& \rho(t) L(t) X,\;  \; X(0)=X_0,
\label{e6101} 
\end{eqnarray}
where $X$ is a vector (i.e., a multiflavor wave function) 
or a matrix (i.e., a density matrix) of arbitrary  dimension.
Equation (\ref{e6101}) is defined in an interaction representation, 
 any additive non-stochastic term has been solved for  and absorbed 
in the definitions of $X(t)$ and $L(t)$.
$L(t)$ is a 
general linear operator.
It can be an ordinary matrix  if $X$ represents a wave function. If $X$ is a density matrix, $L(t)$ is   a commutator:
\begin{equation}
L_A(t) X\equiv [A(t), X].
\end{equation}
Any other linear operator is admissible: obviously it is  
always possible to define
an enlarged vector space where the action of $L$ is represented by a matrix.  

We will assume that 
 $\rho(t)$ in Eq.(\ref{e6101})  
is a scalar Gaussian process completely determined 
by its  first  two moments, which, without loss of generality can be taken as
\begin{eqnarray}
\langle \rho(t)\rangle=0, & & \langle\rho(t)\rho(t')\rangle=f(t,t').
\end{eqnarray}
In other terms, 
$\rho(t)$  is 
characterized completely by the measure 
\begin{eqnarray}
[d\rho] \exp -\frac{1}{2}\int_{-\infty}^\infty \rho(t)f^{-1}(t,t')\rho(t')
 dtdt'. 
\label{e3bb}
\end{eqnarray}

An important particular case is when the process is $\delta$-correlated, 
 the correlation function is then of the form 
\begin{eqnarray}
f(t,t')=\Omega^2 \delta\left (\mid t-t'\mid\right ).
\label{e7001}
\end{eqnarray}
 A  convenient  way of parametrizing a 
 correlation function with a finite 
correlation length is to use an exponential function:
\begin{eqnarray}
f(t,t')=\Omega^2 \epsilon \exp\left (-\epsilon\mid t-t'\mid\right ).
\label{e6105}
\end{eqnarray}

In this case the correlation length is defined as  
$\tau=1/\epsilon$. The expression (\ref{e7001}) is 
reobtained letting  
$\tau\to 0 $ or  $\epsilon\to \infty$ in Eq.(\ref{e6105}).

$X(t)$, the solution to Eq.(\ref{e6101}), is a stochastic function.
The objective of this work is to obtain 
equations for
its ensemble average and higher moments in a systematic way . 
Let us remark that 
for us Eq.(\ref{e6101}) is purely phenomenological, we assume that it is the
result of a more complete microscopic analysis which can account for  the
randomness of $\rho(t)$ (see for example, Refs.\cite{bur1,elm1,otros1,otros2}).

To obtain a differential equation for the average of $X$ we make
use of the following well known property: For any Gaussian process $\rho(t)$ characterized by a $\delta$-correlation function   
(\ref{e6105}) and  any functional $F[\rho]$ 
we have the following relation \cite{zinnj}:
\begin{equation}
\langle F[\rho] \rho(t)\rangle=\Omega^2 \left\langle\frac{\delta F[\rho]}{\delta \rho(t)}\right\rangle.
\end{equation}
For a Gaussian process with an arbitrary correlation function [Eq.~(\ref{e3bb})] 
we have instead the general relation:
\begin{equation}
\langle F[\rho] \rho(t)\rangle=\int d\tau \left\langle\rho(t)\rho(\tau)\right\rangle\left\langle\frac{\delta F[\rho]}{\delta \rho(\tau)}\right\rangle.
\label{e5201}
\end{equation}

 Let us consider 
now the evolution operator $U$  for a particular realization of
Eq.(\ref{e6101}). By definition,
\begin{equation}
X(t)=U(t,t_0) X_0.
\label{e5202}
\end{equation}
The operator $U(t,t_0)$ is a functional of $\rho$, it has  the following formal expression in terms of a time ordered exponential:
\begin{equation}
U(t,t_0)=T \exp\left [ -i \int_{t0}^t\ \rho(\tau) L(\tau) d\tau \right ].
\label{e9003}
\end{equation}
The functional derivative of $U$ with respect  to $\rho(t)$ can be computed by direct methods. By differentiating term by term the series expansion for Eq.(\ref{e9003}), the result is
\begin{equation}
\frac{\delta U(t,t_0)}{\delta \rho(\tau)}=-i L(\tau) U(\tau,t_0), 
\; t<\tau<t_0.
\label{e5203}
\end{equation}

In order to obtain a differential equation for $\avX$ 
we observe that
\begin{equation}
i\langle\partial_t X\rangle=i \partial_t\avX=L(t) \langle\rho(t) X\rangle.
\end{equation}
Combining together Eqs.~(\ref{e5201}),(\ref{e5202}),(\ref{e5203})  
we easily obtain the result 
that  the evolution equation for the ensemble average is in the general case 
an integrodifferential equation given by
\begin{eqnarray}
i\partial_t \avXt&=&-i\int_0^t dt'\ f(t,t') L(t)L(t')\langle X(t')\rangle,
\;  \langle X(0)\rangle =X_0.
\label{e6102}
\label{e9004}
\end{eqnarray}
This equation, which is exact and of very general validity, is the equation 
which
we were looking for and one of the main results of the present work.
Note that previously   
integrodifferential equations have been obtained, 
valid for particular cases or in particular limits, using heuristic 
{\it ad hoc} arguments (for example, in Ref.~\cite{enq3}). The derivation of  Eq.(\ref{e6102}) which has been done here  
is the rigorous justification for such approaches.

In some notable cases Eq.(\ref{e9004}) can be reduced to an ordinary 
differential equation. In the next sections we will see how it can be
reduced to, respectively, first and second order equations for the 
correlation functions given by Eqs.~(\ref{e7001}) and (\ref{e6105}).

\section{THE PARTICULAR  $\delta$-CORRELATED CASE}

For the particular case where the correlation function is of exponential 
type, a second order ordinary differential equation can be obtained as we will see below. 
On the other hand,  
for the simpler $\delta$-correlated case
the evolution equation (\ref{e6102}) becomes the 
ordinary differential equation
\begin{eqnarray}
i\partial_t \avXt&=&-i \Omega^2 L^2(t)\avXt.
\label{e6103}
\label{e9005}
\end{eqnarray}
Taking $L$ as a commutator,
this last equation coincides with the Redfield equation derived by 
Ref.~\cite{lor1}. Note that the effective 
"Hamiltonian" appearing in the second part of Eq.~(\ref{e6103}) is not  
Hermitic anymore (an example of a fluctuation-dissipation effect).

In  cases of interest for the neutrino oscillation problem, the 
original equation for the density matrix  is of 
the slightly simpler form
\begin{eqnarray}
i\partial_t X &=& \left [H_0(t)+\rho(t) g(t) H_1,  X\right ],\;  \; X(0)=X_0,
\label{e6101c} 
\end{eqnarray}
where we have written  the commutator explicitly. $\rho(t)$ is a stochastic function as before and $g(t)$ an arbitrary
scalar function. $H_0,H_1$ are Hamiltonian matrices, the former contains the 
average part of $\rho(t)$:
$H_0\equiv H_0'(t)+\rho_0(t) H_1$, with $\rho_0=\langle\rho\rangle$. 
The latter is assumed to be time independent.

For the problem described by Eq.~(\ref{e6101c})
 the corresponding Redfield equation for the averaged 
density matrix  is of the form
\begin{eqnarray}
i\partial_t \avX
&=& [H_0(t), \avX]-i \Omega^2 g^2(t)[H_1,[H_1,\avX]],\\
&\equiv& H_0^- \avX-\avX H_0^++2 i \Omega^2 g^2(t) H_1 \avX H_1 ,
\label{e6107c}
\end{eqnarray}
where in the last line the following  effective Hamiltonians were defined:
$$H_0^\pm=H_0\pm i g^2(t) H_1^2.$$

The solution of what is called the ``coherent'' part of Eq.~(\ref{e6107c}) 
(two first terms of the Hamiltonian)
is accomplished 
by defining the average evolution operator:
\begin{eqnarray}
\langle U^\pm\rangle&=&T \exp \int d\tau H_0^\pm (\tau), \;\; \left\langle U^-\right\rangle=\left\langle U^{+}\right\rangle^{\dagger}.
\end{eqnarray}
The   coherent part of the density matrix is then
 \begin{eqnarray}
\avX_{coh}&=&\left\langle U^- \right\rangle X_0 \left\langle U^{+}\right\rangle
^{\dagger}.
\label{e4603}
\end{eqnarray}
Defining a new ``coherent'' interaction representation
by the relations 
 \begin{eqnarray}
H_L&=&\langle U^-\rangle^{-1} H_1 \langle U^-\rangle, \;\; 
H_R=\langle U^-\rangle H_1 \langle U^-\rangle^{-1},
\;\;
\avX_I=\langle U^-\rangle \avX_{coh} \langle U^-\rangle^{-1},
\label{e4602}
\end{eqnarray}
we arrive at the equation we were looking for,
the resolution 
of the original equation is equivalent to the resolution 
 of the following one:
 \begin{eqnarray}
i\partial_t \avX_{I}&=&2 i \Omega^2  g^2(t) H_L \avX_I H_R.
\label{e4601}
\end{eqnarray}

There are some  important particular  
cases where  Eq.~(\ref{e9005}) can be
solved or considerably simplified 
by taking into account the algebraic
 properties of a specific  $L$ (in what follows $k(t)$ is  always a 
scalar function).

{(a)} Let us assume that $L$ is such that $L^2(t)=k(t) L(t)$. 
 This case appears in the computation of the average wave function
with matter density noise.
 Equation~(\ref{e9005}) reduces to 
$$i\partial_t \avX=-i \Omega^2 k(t) L(t) \avX.$$
The averaged equation is similar  to the original one, the nonrandom part 
of the density is
``renormalized'' acquiring an imaginary term
$$\rho\rightarrow \rho_0-i \Omega^2 k.$$
This is the density which will appear in the coherent effective Hamiltonians
$H_0^\pm$.

{(b)} The case where $L^2(t)=k(t) I$ with $I$   the identity matrix
 appears in the computation of the averaged neutrino 
wave function under noisy magnetic spin-flavor precession.  
The resulting equation
can trivially be integrated 
(to be compared with the previous case):
$$\langle X(t)\rangle=\exp\left [ -\Omega^2\int_0^t d\tau k(\tau) \right ] \  X_0. $$
The average wave oscillation is damped by a factor equivalent to the one first calculated
by Nicolaidis \cite{nic2}. A similar
 damping also appears  when computing
the average density matrix from Eq.~(\ref{e4601}). 
From these differences of behavior with respect to case (a) it is 
expected  that the presence of 
magnetic field noise can have some influence even if applied far from any 
resonance region.

{\bf (c)} The case where  $L^4(t)=-k(t) L^2(t)$ appears in the 
computation of the
average density matrix with  matter or magnetic noise.
We can obtain in this case the ``conservation law''
$$\left [1-\Omega^2 k(t)\right ]L^2(t) \partial_t \avXt=0.$$
$L^2(t)$ is not invertible because the operator $L(t)$  has a 
zero eigenvalue. The previous expression
has proved to be of practical importance in  
some concrete numerical applications \cite{tor3,tor4,tor5}.

The presence of zero eigenvalues distinguishes cases (a) and (b) from (c); it can 
have consequences in the long term behavior of the respective 
ensemble averages as  is shown elsewhere \cite{tor6}.

\section{AN ASYMPTOTIC EXPANSION FOR EXPONENTIALLY CORRELATED SYSTEMS}
 
We will see now  how  Eq.~(\ref{e9005}) can be obtained 
as a limiting case when the correlation length tends to zero.
For this purpose we use 
an exponential correlation function as Eq.~(\ref{e6105}),  
the integrodifferential 
evolution equation becomes in this case
\begin{eqnarray}
i\partial_t \avXt &=&-i \Omega^2  \ \epsilon\exp(-\epsilon t)\int_0^t 
dt' \exp(\epsilon \tau) L(t)L(t') \langle X(t')\rangle.
\label{e6106}
\end{eqnarray}

Let us   compute the asymptotic expansion of the second term of 
Eq.~(\ref{e6106}) valid for large $\epsilon$; the following 
 expansion is valid  for any function $g(t)$: 
\begin{eqnarray}
h(\epsilon)&\equiv& \epsilon \exp(-\epsilon t)\int_0^t d\tau\ \exp(\epsilon 
\tau) g(\tau)\sim g(t)-\frac{g'(t)}{\epsilon}+\frac{g''(t)}{\epsilon^2}+\cdots
\end{eqnarray}
Inserting this expression into Eq.~(\ref{e6106}), we obtain the following 
expansion in 
powers of $\epsilon$:
\begin{eqnarray}
i\partial_t \avX&=&-i \Omega^2 L^2(t)\langle X\rangle+i \frac{\Omega^2}{\epsilon} L(t) 
\partial_t\left ( L(t)\avX\right )+o\left(\frac{1}{\epsilon^2}\right ).
\end{eqnarray}
To leading order in $1/\epsilon$, we recover the expression corresponding to 
the $\delta$-correlated case. At next-to-leading order we get  
finite-correlation correction terms 
\begin{eqnarray}
i\partial_t \avX&=&-i \Omega^2 L^2(t)\avX+i\frac{\Omega^2}{\epsilon}L(t)L'(t)
 \avX+i 
\frac{\Omega^2}{\epsilon}L^2(t)\partial_t \avX
\end{eqnarray}
or, equivalently,
\begin{eqnarray}
\left[ 1-\frac{\Omega^2}{\epsilon} L^2(t)\right ]\partial_t \avX=\left [-\Omega^2 
L^2(t)+\frac{\Omega^2}{\epsilon}L(t)L'(t)\right ] \avX.
\end{eqnarray}
Finally,  We get 
the following differential equation 
valid to order $1/\epsilon$, 
 making the assumption that the operator which multiplies 
the left term is invertible: 
\begin{eqnarray}
\partial_t \avX&=&\left (-\Omega^2 
L^2(t)+\frac{\Omega^2}{\epsilon}L(t)L'(t)+\frac{\Omega^4}{\epsilon}L^4(t)\right 
) \avX.
\label{e5205}
\end{eqnarray}

This equation can be used for finite, but relatively large, 
correlation lengths. We see that, up to this degree of approximation, not only 
the ratio level of noise to correlation length ($\Omega^2/\epsilon$) is 
important. We have different regimes 
according to the value of $\Omega^2$. 
The first term will 
be more important for a low noise amplitude ($\Omega^2<<1$). 
For strong noise ($\Omega^2>>1$) 
the second term, proportional 
in this case to $L^4$, will dominate.

\section{AN EXACT DIFFERENTIAL EQUATION FOR EXPONENTIALLY CORRELATED SYSTEMS}
 
In contrast with the 
approximate approach used in the previous section, we can actually  
derive  a simple, ordinary second order differential equation for the case of exponential 
correlation.  
Let us assume that the original equation is of the same decomposable type as
the one appearing in Eq.~(\ref{e6101c}) but let us include other cases using the 
general notation
\begin{eqnarray}
i\partial_t X &=& \left (L_0(t)+\rho(t) g(t) L_1 \right )X,\;  \; X(0)=X_0,
\label{e6101b} 
\end{eqnarray}
with $L_0,L_1$ general linear operators as before, the latter 
time independent.
In this case Eq.~(\ref{e6102}) is of the form
\begin{eqnarray}
i\partial_t \avX= L_0(t) \avX-i \Omega^2 \epsilon L_1^2 e^{-\epsilon t} g(t)\int_0^t dt' g(t') e^{\epsilon t'} \langle X(t')\rangle.
\label{e6107}
\end{eqnarray}
Differentiating the equation once and performing 
 some simple algebra we obtain the following ordinary 
second order differential equation:
\begin{eqnarray}
\left [\partial_t-\lambda(t)\right ]\left [i \partial_t -L_0(t) \right ] \avX&=&
-i \Omega^2 \epsilon g^2(t) L_1^2 \avX,\nonumber\\ 
 i(\partial_t \avX)_0&=&L(0) \avX_0, \;\; \avX_0=X_0,
\label{e9201} 
\end{eqnarray}
with 
$$\lambda(t)\equiv -\epsilon+g'(t)/g(t).$$
Let us remark  that  this equation is exact and probably the most important 
result of this work from a practical point of view.

\section{HIGHER ORDER MOMENTS AND THE QUANTUM REGRESSION THEOREM}
 
Second order  distribution moments, expressions of the type
$\langle X_i X_j\rangle$, or in general, moments of any order, 
can also   be computed  using equations similar to  Eq.~(\ref{e6102}). 
The straightforward procedure is to  
 define products $X_{ij\cdots k}=X_i X_j\cdots X_k$ and write differential equations 
for them using the constitutive equations for each of the $X_i$. 
 The resulting equations are of the same type as  Eq.~(\ref{e6101}).  
The similarity is   obvious when 
one adopts a tensorial notation and defines products of the form
$X^{(n)}=X\otimes\cdots\otimes X. $ The task of obtaining  evolution  equations is especially simple within this notation.

Correlators of quantities at different
times also appear   in the computation of  averages of  
quantities of physical interest, i.e., expected signal rates.  
 In the most simple case expressions 
of the type $\langle X_i(t+\tau) X_j(t)\rangle$.
We will shortly show  that the knowledge of equal-time moments  and the application of the quantum correlation theorem is sufficient for 
the computation of this kind of correlators.

Our objective now is  the computation of the 
correlation function appearing 
in two steps. First 
we define the generalized density matrix $X^{(2)}$ as the tensorial product of 
usual density matrices at two different times and energies:
\begin{eqnarray}
X^{(2)}(E_1,E_2; t_1,t_2)&\equiv& X(E_1,t_1)\otimes X(E_2,t_2).
\label{e6111}
\end{eqnarray}
The  average of the 
element $\langle X^{(2)}_{1111}\rangle=\langle X_{11}X_{11}\rangle$ 
is evidently the probability correlation function we are looking 
for.

The differential equation for the equal time function 
$X^{(2)}(E_1,E_2; t,t)$ is
obtained
 from the individual evolution equations for the matrices 
$X_{1,2}\equiv X(E_{1,2}) $  
[indices (1,2) label expressions where $E_1,E_2$ appear respectively]
\begin{eqnarray}
\partial_t X^{(2)}&=& H_1 X^{(2)}+X^{(2)} H_2\equiv\rho L X^{(2)}. 
\label{e6112}
\end{eqnarray}
Equation~(\ref{e6112}) is a random linear differential equation, linear in the 
stochastic variable $\rho(t)$. 
Applying the formalism developed in the previous 
section, we can immediately write  the equation for the
ensemble average  $\langle X^{(2)}\rangle$ [Eqs.~(\ref{e6102}--\ref{e6103})].
Once we know the equal time correlator, we
obtain the expression for any other pair $(t,t')$ using the 
quantum regression theorem \cite{gard1}
which reads as follows.
For any vector Markov process $Y$, if the ensemble average of $Y$ fulfills
a Schr\"{o}dinger-like equation of the type
\begin{eqnarray}
\partial_t \langle Y(t)\rangle=G(t) \langle Y(t)\rangle,
\end{eqnarray}
with $G(t)$ an arbitrary matrix, then 
the second order correlations will obey the 
following equation:  
\begin{eqnarray}
\partial_\tau \langle Y_i(t+\tau) Y_l(t)\rangle=\sum_j G_{ij}(\tau) \langle Y_j(t+\tau) Y_l(t)\rangle.
\end{eqnarray}

Note that 
the quantum regression theorem is in principle not applicable 
to systems described by the integral equation Eq.~(\ref{e6102}). Nevertheless
 it is 
applicable to  problems where the integral equation can be reduced 
to an ordinary differential equation such as Eq.~(\ref{e9201}). Such a second 
order equation can easily be   expressed as a first order one  
 by defining  the auxiliary pair process $Y=(X,\ \partial_t X)$.  

For the simplest case where $\rho$ is a $\delta$ correlated process
the equation for non-equal time correlators is explicitly given by
\begin{eqnarray}
i \partial_\tau \langle X^{(2)}(t+\tau,t)\rangle&=& -i \Omega^2 L^2(\tau)
\langle X^{(2)}(t+\tau,t)\rangle. 
\label{e6114}
\end{eqnarray}
The initial condition,  
$X^{(2)}(t,t)$, is the equal time correlator previously obtained.
For exponentially correlated problems second order  equations similar to 
Eq.~(\ref{e9201}) can  immediately be obtained.

\section{CONCLUSIONS AND FINAL REMARKS}

In conclusion, in this work 
we have  developed the general 
 formalism for the study of neutrino propagation in 
stochastic media. This formalism has allowed 
 the  systematic derivation of   evolution equations for averaged 
quantities as survival probabilities and higher order moments of them. 
New integrodifferential equations
 valid for finite 
correlation processes have been  obtained for the first time. 
For exponentially correlated processes a second order ordinary equation is 
obtained as a consequence.
The Redfield equation valid
 for Gaussian $\delta$-correlated
noise is rederived in  a simple way: 
it has been obtained directly  and as 
the zero-order term of an asymptotic expansion in 
 the inverse  of the correlation length.

The formalism can be generalized in an obvious way to 
obtain 
  the  ensemble average  of equations of  a  slightly more general type
 than Eq.~(\ref{e6101}), equations of the form
\begin{eqnarray}
i\partial_t X &=& \sum_i\rho_i(t) L_i(t) X,\; \; X(0)=X_0.
\label{e4101} 
\end{eqnarray}
These equations  appear, for example, in the 
chaotic neutrino magnetic precession.
The result for the case where the $\rho_i$ are $\delta$-correlated in time but  mutually uncorrelated is simply 
\begin{eqnarray}
i\partial_t \avXt&=&-i \sum_i\Omega_i^2 L_i^2(t)\avXt.
\end{eqnarray}

The generalization to continuous Schr\"{o}dinger equations in the presence of 
random potentials or random external forces is also obvious if we leave
 aside 
the mathematical differences coming from the appearance of 
infinite-dimensional Hilbert spaces.
The formalism is of general application to any 
quantum system governed by similar Schr\"{o}dinger equations: equations where
a stochastic function appears multiplicatively  
in some term of the Hamiltonian.

\vspace{0.3cm}



\end{document}